\documentclass[traditabstract]{aa}
\usepackage{natbib} 
\usepackage{graphicx}
\usepackage{txfonts}

\begin{document}

\title{Evidence of a double--double morphology in B\,0818+214}

\author{A. Marecki \and M. Szablewski} \offprints{Andrzej Marecki 
\email{amr@astro.uni.torun.pl}} \institute{Toru\'n Centre for Astronomy, N. 
Copernicus University,
           87-100 Toru\'n, Poland}

\date{Received 24 July 2009 / Accepted 19 September 2009}

\abstract{The so-called double--double structure in radio sources is the 
most conspicuous signature of their restarted activity. Observations 
indicate that in the majority of double--double radio sources (DDRS), the 
span of the radio lobes is larger than 0.7\,Mpc. This lower limit is also 
suggested by theory. However, it seemed likely that the apparent core 
of B\,0818+214, a radio galaxy with an overall linear size of 
its radio structure below that limit, could harbour a compact double well 
aligned with the outer lobes so that the whole object would fulfil the 
criteria of a DDRS. Here, we present evidence that the central 
component of B\,0818+214, when magnified through the EVN+MERLIN 18-cm 
observations, shows two FR\,II-like lobes. As the separation of the 
inner lobes is not greater than 5.7\,kpc, they are immersed in the ISM of 
the host galaxy. This circumstance is the likely reason why the inner double 
has become visible, despite the predictions of the theory according 
to which B\,0818+214 as a whole is too small for a new double to develop 
inside the cocoon inflated during the previous active phase. Moreover, we 
speculate that its host galaxy is not active at the moment and so the inner 
double may be in the coasting phase often observed in other medium-sized 
symmetric objects with intermittent activity. It could be, therefore, that 
two different mechanisms of accretion disk instabilities, ionisation and 
radiation-pressure driven, may be independently responsible for triggering 
active phases, manifesting as the outer and the inner doubles, respectively.
}


\maketitle


\section{Introduction}

According to a well-established paradigm, the activity of galaxies can be 
recurrent. For radio-loud active galaxies, \citet{best07} suggested that
their activity {\em must} be constantly re-triggered. Given that the lobes
that build up during the active phase are huge reservoirs of energy -- 
the minimum energy stored in the lobes is in the range of $10^{60}$ to 
$10^{64}$\,ergs \citep[see e.g.][]{rich98} -- it can take up to 
$10^8$\,years for them to disappear once the energy supply from the active 
nucleus cuts off \citep{kg94, slee01}. Because the lifetimes of relic lobes 
are that long, it is plausible that a new active phase of a galaxy may begin 
before the old lobes have faded out completely, so that its radio images show 
components resulting from both current and past active periods. The 
signature of such a renewed activity is most convincing if a large, 
double-lobed relic structure straddles a pair of young lobes giving rise to 
the so-called double-double radio source (DDRS) \citep{lara99, schoen00a, 
schoen00b, sarip02, sarip03, saikia06} or even triple double 
\citep{brocksopp07}. DDRSs are almost entirely identified with 
galaxies; an exception -- quasar J0935+0204 (4C\,02.27) -- has been found 
by \citet{jamrozy09}.

Typically, the separation of the outer lobes in DDRSs is not greater than 
one order of magnitude that of the inner lobes. For example, the 
outer/inner double size ratio for the sources shown in the seminal paper on 
DDRSs by \citet{schoen00a} ranges from 2 (\object{B\,1240+389}) to 9 
(\object{B\,1450+333}). At first sight, this appears puzzling since 
the new lobes/hotspots started to expand much later than the old ones, so 
that they should be much closer to one another. However, hotspots pushed
by restarted jets propagate much 
faster because of the low density of plasma inside the cocoon inflated 
during the previous phase of activity, compared to those of the old 
jets moving through the much denser intergalactic medium (IGM) 
\citep[][hereafter KSR]{kaiser00}. Therefore, even if the length of the 
previous active phase had been the maximum possible ($10^8$\,yr) and assuming 
that the activity then ceased for a substantial amount of time (also of 
the order of $10^8$\,yr), once it was re-triggered, the inner structure
developed quickly enough so that it is able to catch up with the outer 
one. It is speculated -- see e.g. KSR -- that the inner lobes 
can reach the old cocoon boundary, cross it, and then overtake the outer 
ones. Consequently, these objects could, paradoxically, also be treated as 
DDRSs, or ``inverted DDRS''.
\object{PKS\,0349$-$27} seems to be an extreme case of such a scenario -- 
see the image presented by \citet{morganti93}. Here, the new lobes have
overtaken the old ones and appear as completely separate entities. As a 
result, the whole source is clearly double-double but, unlike in standard 
DDRSs, in PKS\,0349$-$27, the relics are not the outer but the inner pair.

Because of that rapid development of the inner lobes, it would appear that 
DDRSs where the separation of the inner lobes is significantly smaller than 
that of the outer lobes should be rare. Yet, \object{3C\,236} \citep{sch01} 
and \object{J1247+6723} \citep{marecki03} are such special cases of 
double-double morphology where the inner part is too compact to be properly 
imaged in the maps encompassing the outer one and so, as a whole, the source 
does not appear as a DDRS but as a core-dominated triple (CDT), where 
the alleged core is actually a compact, luminous double source. J1247+6723 
is the most extreme case of such situation known to date; here, the ratio of 
the size of the outer to inner double is five orders of magnitude. 
These two examples are clear indication that at least some CDTs could 
be concealed DDRSs. \citet{konar04} pointed out that the central 
component of a CDT was more likely to be a compact double if that component 
had a steep or gigahertz-peaked spectrum (GPS)\footnote{In fact, the inner 
structure of J1247+6723 is a GPS source.}. Hence, investigations of CDTs 
with GPS or steep-spectrum cores could lead to discoveries of more DDRSs 
where a new active phase was triggered only recently so that the linear 
sizes of the inner doubles are still small.

The catalogue of radio sources extracted from the {\em Faint Images of Radio
Sky at Twenty centimetres} (FIRST) survey \citep{white97}\footnote{Official 
website: http://sundog.stsci.edu} is very well suited for machine-aided 
search for CDTs. To this end, \citet{marecki06} -- hereafter Paper\,I -- 
developed an automated procedure. They also used the GB6 catalogue \citep{GB6} 
to calculate spectral indices between the frequencies of these two surveys 
and selected only steep-spectrum sources, i.e. those where $\alpha > 0.5$ 
($S\propto\nu^{-\alpha}$). Obviously, calculation of the spectral indices of 
the ``cores'' alone was not possible because of the resolution limit of the 
GB6 survey so this criterion was fulfilled only approximately. 
A large fraction of core-jet sources, which typically have flat spectra, 
possibly were eliminated in this way.

A sample of 15 sources resulted from this selection procedure and 
is given in Table\,1 of Paper\,I. Their cores were 
observed with MERLIN at 6\,cm -- see Paper\,I. As a result, 
\object{B\,0818+214} emerged as the most likely candidate for a DDRS since 
it appeared to have two components in the centre very well 
aligned with the outer lobes (Paper\,I, Fig.\,1). Admittedly, the inner 
components were largely asymmetric so, at that stage, a core-jet morphology 
could not be ruled out. Whether the inner double was actually two 
FR\,II-like \citep{fr74} lobes could only be decided by means of a 
low-frequency, high-resolution observation. Thus, B\,0818+214 was followed 
up at 18\,cm (1658\,MHz) with the European VLBI Network (EVN) supplemented 
with MERLIN. Here, the outcome of these latest observations is reported.


\section{Observations and results}

The overall structure of B\,0818+214 as seen in FIRST image is shown in 
Fig.\,1 of Paper\,I. The source has three aligned components, the middle 
one being roughly an order of magnitude stronger than the remaining 
two -- see Table\,2 of Paper\,I. The dominance of the central component is 
particularly obvious when the peak flux densities are compared:
6-7\,mJy for each of the outer lobes and 142\,mJy for the ``core''. 
The peak flux densities of the lobes are about 3\,times less 
than their total flux densities, which may suggest that these 
structures are diffuse as expected for relic lobes.

High-resolution observations of the ``core'' of B\,0818+214 were carried out 
on 10~June~2007 at 1658\,MHz with phase referencing. The network consisted 
of 14 telescopes of the EVN and MERLIN including the largest two, Lovell and 
Effelsberg, although due to technical limitations of the combined 
EVN+MERLIN network, visibilities for only 41 baselines were obtained. 
Nevertheless, this and the length of the total time spent on the program 
source -- nearly 4\,hours divided into 23 10-min. scans spread over 
7.5\,hours -- resulted in a very good, uniform $u$--$v$ coverage and good 
sensitivity: the noise in the resulting image was $\sim$40\,$\mu$Jy/beam.

\begin{figure}[ht]
\includegraphics[scale=0.4]{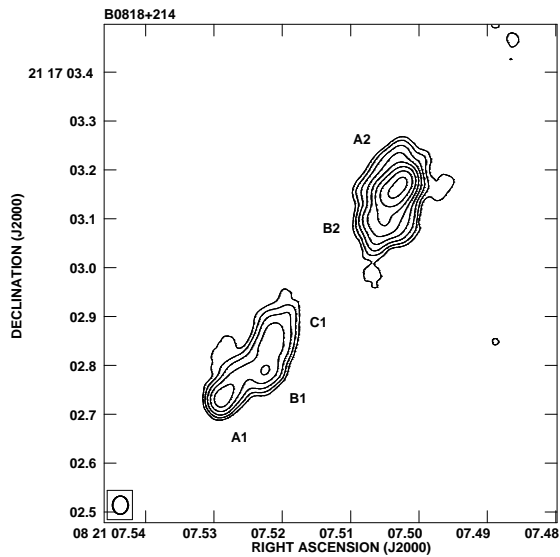} 
\caption{EVN+MERLIN image of the inner double of B\,0818+214 at 1658\,MHz.
Contours increase by a factor of 2; the first contour level corresponds to
$\approx 5\sigma$ level which is 0.21\,mJy/beam. The peak flux density is
35.1\,mJy/beam. The beam size is $37 \times 32$\,milliarcseconds at the
position angle of $0\degr$.} 
\label{fig:evn}
\end{figure}

The reduction of the data using AIPS was carried out in the standard way and 
produced the image shown in Fig.~\ref{fig:evn}.

The 18-cm flux densities of the components of the southeastern lobe (A1, B1, 
and C1) were measured by fitting Gaussian models using the 
AIPS task JMFIT and are listed in column\,2 of Table~\ref{comp_data}. 
Column\,3 gives the flux densities of their equivalents at 6\,cm (4994\,MHz) 
extracted from the MERLIN image shown in Fig.\,1 of Paper\,I. Due to the
limited resolution of the MERLIN 6-cm map, it was not possible to fit a 
Gaussian model for component B2 of the northwestern lobe in that image. 
Thus, we used the AIPS TVSTAT utility to measure the total flux densities of 
that lobe for both 18-cm and 6-cm images. They are given in the bottom 
row of Table~\ref{comp_data}.

\begin{table}[ht]
\caption[]{Flux densities and spectral indices of the 
components of the inner double.}
\centering
\begin{tabular}{c r r c}
\hline
\hline
Component & \multicolumn{2}{c} {Flux density [mJy]} & Spectral index\\
& 1658\,MHz & 4994\,MHz & \\
\hline
A1 &     11.7 &  6.2 & 0.57 \\
B1 &      9.9 &  4.4 & 0.75 \\
C1 &     10.9 &  2.3 & 1.40 \\
A2+B2 & 108.6 & 39.1 & 0.93 \\
\hline
\end{tabular}
\label{comp_data}
\end{table}

Based on the flux density figures in columns 2 and 3, the spectral indices 
of the source's inner structure have been calculated and are shown in
column\,4. The values of the indices make it clear that A1, having the
flattest spectrum, must be a hotspot. Since the spectrum becomes steeper to
the north-west, indicating increased spectral age in that 
direction, the whole southeastern feature of the inner structure appears as 
a typical FR\,II lobe. The morphology and steep spectrum of the
A2--B2 region also leaves no doubt that it is the second lobe.


\section{Discussion}

KSR developed a general theory of DDRSs. The main conclusion of their 
work is that the interior of the old cocoon is an environment where the lobes
resulting from a new phase of the nuclear activity could not develop 
easily. For example, KSR show that the scenario of a mere replacement of the 
cocoon material by the surrounding IGM is inefficient as it would 
require $10^8$\,yr to refill the cocoon once the bow shock has disappeared. 
Since the above replacement time is close to the decay time of the lobes and 
disappearance of the bow shock can also take a considerable amount of 
additional time, DDRSs should rarely be observed as even the largest/oldest 
ones would never be mature enough (in terms of the internal density) to host 
a new double.

In an alternative model, KSR posit the presence of warm clouds of gas embedded 
in the hot IGM. Unlike in the previous scenario, such clouds are able to pass 
the cocoon border even if there is still a bow shock surrounding the outer 
source structure. Having done so, they break up into smaller fragments 
inside the cocoon. This dispersion process takes $\sim$$5\times10^7$\,yr. 
Such a timescale, albeit of comparable magnitude, gives the chance 
for new lobes and hotspots to develop but only in radio objects with linear 
sizes of more than $\approx$700\,kpc. It follows that, even if the 
activity in a younger source could actually stop and then restart, such an 
event would remain unobservable.

B\,0818+214 lies within the Sloan Digital Sky Survey footprint and it is 
listed there as an $m_R=21.73$ galaxy. Since no redshift is 
available, only the upper limits to the linear sizes of both the outer and 
the inner doubles can be established. Given that for the up-to-date 
cosmological parameters \citep[$H_0=71$\,km~s$^{-1}$/Mpc, $\Omega_M=0.27$, 
$\Omega_\Lambda=0.73$,][]{wmap03} the maximum possible angular-size distance 
$D_{A max}=1765.1$\,Mpc (for $z=1.64$), the largest ratio of linear to 
angular sizes for any object in the universe is 8.558\,kpc$/\arcsec$. Hence, 
the overall angular size of B\,0818+214, $68\arcsec$, yields its maximum 
linear size of 582\,kpc whereas the angular size of the inner double,
$0\farcs67$, translates to not more than 5.7\,kpc. These are the 
projected sizes, but the absolute sizes are not likely to be significantly 
greater. This is because if the angle $\theta$ between the source axis and 
the line of sight was considerably less than $90\degr$ then the light 
travel-time difference between the epochs in which we observe the outer 
lobes of $1.9/\tan\theta$\,Myr would be of the order of 
$10^6$\,yr. Such a large time lag would inevitably result in a noticeable 
asymmetry of the outer lobes seen at two well-separated stages of the decay.
This is not observed for the outer lobes of B\,0818+214: their flux densities,
armlength ratios, and shapes are very similar.

As the upper limit for the linear sizes of B\,0818+214 is below the 
threshold required by KSR, it might appear that their model is not 
applicable to radio sources like B\,0818+214. However, the problem 
can be resolved taking into account that B\,0818+214 is not a typical DDRS 
since the inner lobes are two orders of magnitude more compact than the 
outer ones. Hence, although the inner double is not as small as that of 
J1247+6723 where we observe an FR\,II-like double with the linear size 
of the order of 10\,pc \citep{marecki03}, here, we have a kpc-scale, i.e. 
medium-sized symmetric object \citep[MSO,][]{fanti95}. Because of its 
size, it is not immersed in the low-density medium of the outer cocoon as 
in standard DDRSs but rather lies in the ISM of the host galaxy and strongly
interacts with it. Such interactions are observed in many compact sources --
see \citet{labiano05, labiano06}, \citet{holt06}, \citet{labiano08} -- and 
should be expected in B\,0818+214. This circumstance explains why it could 
have developed at all and become observable despite the predictions of KSR.

Another potential problem is that the inner lobes of B\,0818+214 are 
highly asymmetric in terms of their flux densities but, again, this is quite 
normal as, due to interactions with asymmetric medium in the central regions 
of their host galaxies, compact steep spectrum (CSS) sources are more 
asymmetric than larger sources of similar powers \citep{jey05}. The 
majority of MSOs are CSSs and this is the case for the inner part of 
B\,0818+214. Another factor that may contribute to its observed 
asymmetry is the orientation. If the inner lobes have just entered the 
coasting phase -- see the next paragraph -- the fast decay of such compact 
lobes combined with the light travel-time effect could be responsible for 
the differences in their appearances.

There is yet another characteristic that makes B\,0818+214 a special object. 
The majority of well-defined DDRSs, apart from two pairs of lobes, have a 
core. While this is the case for \object{B\,0925+420}, \object{B\,1240+389}, 
\object{B\,1450+333} \citep{schoen00a}, \object{B\,1834+620} 
\citep{schoen00b}, \object{J0116$-$473} \citep{sarip02}, and 
\object{B\,1545$-$321} \citep{sarip03}, we have found no core in B\,0818+214.
This could mean that the energy transport from the core has ceased, 
although, taking into account that at least one hotspot is still visible, 
this should have happened very recently. Given that the linear size of the 
MSO in the centre of B\,0818+214 is a few kpc and assuming a lobe 
expansion rate of the order of $0.1c$ \citep{scheuer95}, it appears that the 
cessation of activity happened after a relatively short active period with 
a length of the order of $10^5$\,yr. The duration of the active period 
may be even shorter because lobe expansion rates for DDRSs could 
equally well be even $0.3c$ as shown by \citet{schoen00b} for B1834+620.  
Therefore, the inner part of B\,0818+214, like other MSOs, could be 
intermittent on a timescale of $10^4 - 10^5$\,yr as originally theorised by 
\citet{rb97}. Recently, a short-timescale intermittency in young radio 
sources resulting from radiation-pressure-driven instabilities in accretion 
disks was modelled by \citet{czerny09} and studied on samples of GPS/CSS 
sources by \citet{wu09a, wu09b}.

The assumption that the MSO in the centre of B\,0818+214 is in the coasting 
phase leads to the conclusion that the behaviour of the active nucleus in
B\,0818+214 fits two different evolutionary tracks. On the one 
hand, its outer lobes resulted from a long phase of activity interrupted due 
to e.g. ionisation instability operating on longer timescales \citep[][and 
references therein]{hse01, janiuk04}. On the other hand, in the most recent 
activity cycle, the central source is intermittent due to radiation-pressure 
instability operating on much shorter timescales \citep{czerny09}. Hence, it 
appears that two accretion disk instability mechanisms, ionisation and 
radiation-pressure driven, could be at work in the same galaxy and the 
outburst resulting from the short-timescale instability may occur during the 
quiescent phase of the long-timescale instability. A direct consequence 
of this scenario is that B\,0818+214 in its present form would not evolve to 
a standard DDRS.

\section{Concluding remarks}

Recurrent activity in galaxies is likely to be much more ubiqui\-tous than 
it appears and the re-triggering timescales can span several orders of 
magnitude. The main reason for why we do not see many such events in the
form of a DDRS is that young jets often travel silently in the cocoon 
inflated during the previous active phase because of its low density and so 
they do not develop shocks/hotspots. On the other hand, the quiescent state 
of an active nucleus for a given cycle can be longer than the decay time of
the outer lobes. This could explain the negative result of the search for
relics carried out by \citet{sirothia09}. Therefore, it looks as if some 
fine-tuning between the lifetime of the outer coasting structures and the 
time needed to refill the cocoon is required to make it possible for a 
standard DDRS phenomenon to take place and that fine-tuning often 
does not work.

Out of the wide spectrum of sizes of the inner doubles, only radio sources
with either the largest or the smallest ones can be observed: the former are 
perceived as classical DDRSs, whereas the latter appear as CDTs and only 
high-resolution observations of their ``cores'' can reveal their true 
nature. For the latter group of objects, a sub-galactic size of the inner 
double is likely so that the radio source is embedded in the host galaxy. 
This circumstance makes the above-mentioned fine-tuning not necessary for 
them as the ISM is always dense enough for an FR\,II-like structure to 
develop and so whether the outer cocoon has already been refilled or not is 
irrelevant. CDTs are, therefore, a potential reservoir of numerous 
DDRSs with the caveat that {\em not every} CDT will turn out to 
be a DDRS. For example, \citet{marecki04} presented a 
high-resolution observation of a CDT source \object{J1708+0035} whose 
central component is not a compact double but a core-jet structure. 
Moreover, a conspicuous misalignment between the axes of the inner and the 
outer structures is present in that source which makes it belong to a 
different class of object. Further specimens of this class are shown and an
interpretation is given in Paper\,I.

\begin{acknowledgements}

\item The European VLBI Network is a joint facility of European, Chinese, 
South African and other radio astronomy institutes funded by their national 
research councils.

\item MERLIN is operated by the University of Manchester as a National 
Facility on behalf of the Particle Physics \& Astronomy Research Council 
(PPARC).

\item AM acknowledges the receipt of a travel grant funded by FP6 Radio\-Net 
as a part of the Trans-National Access (TNA) programmes.

\end{acknowledgements}


\begin{thebibliography}{}

\bibitem[Becker et al.(1991)]{GB6} Becker, R.~H., White, R.~L., \& Edwards, A.~L.\ 1991, \apjs, 75, 1
\bibitem[Best(2007)]{best07} Best, P.~N.\ 2007, New Astronomy Review, 51, 168
\bibitem[Brocksopp et al.(2007)]{brocksopp07} Brocksopp, C., Kaiser, C.~R., Schoenmakers, A.~P., \& de Bruyn, A.~G.\ 2007, \mnras, 382, 1019
\bibitem[Czerny et al.(2009)]{czerny09} Czerny, B., Siemiginowska, A., Janiuk, A., Nikiel-Wroczy{\'n}ski, B., \& Stawarz, {\L}.\ 2009, \apj, 698, 840
\bibitem[Fanaroff \& Riley(1974)]{fr74} Fanaroff, B.~L., \& Riley, J.~M.\ 1974, \mnras, 167, 31P
\bibitem[Fanti et al.(1995)]{fanti95} Fanti, C., Fanti, R., Dallacasa, D., et~al.\ 1995, \aap, 302, 317
\bibitem[Hatziminaoglou et al.(2001)]{hse01} Hatziminaoglou, E., Siemiginowska, A., \& Elvis, M.\ 2001, \apj, 547, 90
\bibitem[Holt et al.(2006)]{holt06} Holt, J., Tadhunter, C.~N., \& Morganti, R.\ 2006, Astronomische Nachrichten, 327, 147
\bibitem[Jamrozy et al.(2009)]{jamrozy09} Jamrozy, M., Saikia, D.~J., \& Konar, C. \mnras~accepted, arXiv:0908.1508
\bibitem[Janiuk et al.(2004)]{janiuk04} Janiuk, A., Czerny, B., Siemiginowska, A., \& Szczerba, R.\ 2004, \apj, 602, 595
\bibitem[Jeyakumar et al.(2005)]{jey05} Jeyakumar, S., Wiita, P.~J., Saikia, D.~J., \& Hooda, J.~S.\ 2005, \aap, 432, 823
\bibitem[Kaiser et~al.(2000)]{kaiser00} Kaiser, C.~R., Schoenmakers, A.~P., \& R\"ottgering, H.~J.~A.\ 2000, \mnras, 315, 381 (KSR)
\bibitem[Komissarov \& Gubanov(1994)]{kg94} Komissarov, S.~S., \& Gubanov, A.~G.\ 1994, \aap, 285, 27
\bibitem[Konar et al.(2004)]{konar04} Konar, C., Saikia, D.~J., Ishwara-Chandra, C.~H., \& Kulkarni, V.~K.\ 2004, \mnras, 355, 845
\bibitem[Labiano et al.(2005)]{labiano05} Labiano, A., et al.\ 2005, \aap, 436, 493
\bibitem[Labiano et al.(2006)]{labiano06} Labiano, A., et al.\ 2006, \aap, 447, 481
\bibitem[Labiano(2008)]{labiano08} Labiano, A.\ 2008, \aap, 488, L59
\bibitem[Lara et al.(1999)]{lara99} Lara, L., M{\'a}rquez, I., Cotton, W.~D., et~al.\ 1999, \aap, 348, 699
\bibitem[Marecki et al.(2003)]{marecki03} Marecki, A., Barthel, P.~D., Polatidis, A., \& Owsianik, I.\ 2003, PASA , 20, 16
\bibitem[Marecki(2004)]{marecki04} Marecki, A.\ 2004, Proceedings of the 7th European VLBI Network Symposium, 12-15 October 2004, Toledo, ed. R. Bachiller et al., 117
\bibitem[Marecki et al.(2006)]{marecki06} Marecki, A., Thomasson, P., Mack, K.-H., \& Kunert-Bajraszewska, M.\ 2006, \aap, 448, 479 (Paper\,I)
\bibitem[Morganti et al.(1993)]{morganti93} Morganti, R., Killeen, N.~E.~B., \& Tadhunter, C.~N.\ 1993, \mnras, 263, 1023
\bibitem[Reynolds \& Begelman(1997)]{rb97} Reynolds, C.~S., \& Begelman, M.~C.\ 1997, \apjl, 487, L135
\bibitem[Richstone et al.(1998)]{rich98} Richstone, D., Ajhar, E.~A., Bender, R., et~al.\ 1998, \nat, 395, A14
\bibitem[Saikia et al.(2006)]{saikia06} Saikia, D.~J., Konar, C., \& Kulkarni, V.~K.\ 2006, \mnras, 366, 1391
\bibitem[Saripalli et al.(2002)]{sarip02} Saripalli, L., Subrahmanyan, R., \& Udaya Shankar, N.\ 2002, \apj, 565, 256
\bibitem[Saripalli et al.(2003)]{sarip03} Saripalli, L., Subrahmanyan, R., \& Udaya Shankar, N.\ 2003, \apj, 590, 181
\bibitem[Schilizzi et~al.(2001)]{sch01} Schilizzi, R.~T., Tian, W.~W., Conway, J.~E., et~al.\ 2001, \aap, 368, 398
\bibitem[Schoen\-makers et al.(2000a)]{schoen00a} Schoenmakers, A.~P., Bruyn, A.~G., R\"ottgering, H.~J.~A., van der Laan, H., \& Kaiser, C.~R.\ 2000a, \mnras, 315, 371
\bibitem[Schoen\-makers et al.(2000b)]{schoen00b} Schoenmakers, A.~P., Bruyn, A.~G., R\"ottgering, H.~J.~A., van der Laan, H.\ 2000b, \mnras, 315, 395
\bibitem[Scheuer(1995)]{scheuer95} Scheuer, P.~A.~G.\ 1995, \mnras, 277, 331
\bibitem[Sirothia et al.(2009)]{sirothia09} Sirothia, S.~K., Saikia, D.~J., Ishwara-Chandra, C.~H., \& Kantharia, N.~G.\ 2009, \mnras, 392, 1403
\bibitem[Slee et al.(2001)]{slee01} Slee, O.~B., Roy, A.~L., Murgia, M., Andernach, H., \& Ehle, M.\ 2001, \aj, 122, 1172
\bibitem[Spergel et al.(2003)]{wmap03} Spergel, D.~N., et al.\ 2003, \apjs, 148, 175
\bibitem[White et al.(1997)]{white97} White, R.~L., Becker, R.~H., Helfand, D.~J., \& Gregg, M.~D.\ 1997, \apj, 475, 479
\bibitem[Wu(2009a)]{wu09a} Wu, Q.\ 2009a, \mnras, 398, 1365
\bibitem[Wu(2009b)]{wu09b} Wu, Q.\ 2009b, \apjl, 701, L95

\end{thebibliography}
\end{document}